\documentclass[twocolumn,showpacs,preprintnumbers,amsmath,amssymb,aps]{revtex4}

\usepackage{verbatim} 
\usepackage{graphicx}
\usepackage{dcolumn}
\usepackage{bm}

\def\N{{\cal N}}

\def\S_1{{\widetilde {S_1}}}

\def\R{{\mathbb R}}

\def\tr{{\rm tr}}

\def\Dslash{{\rlap{\raise 1pt \hbox{$\>/$}}D}}

\begin{document}

\preprint{SLAC-PUB-13285}

\title{Quantum phase transitions  and new scales in  QCD-like theories }

\author{Mithat \"Unsal}%
\affiliation{%
SLAC and Department of Physics,  Stanford University, CA, 94309/94305   \\
}%

\date{\today}

\begin{abstract}
It is commonly believed that in confining vector-like gauge theories the center and   chiral symmetry realizations are  parametrically entangled, and   if  phase transitions occur,   they 
 must take place around    the  strong scale $\Lambda^{-1}$  of the gauge theory. 
We demonstrate that (non-thermal) vector-like theories   formulated on 
${\mathbb R}^{3} \times S^1$ 
where $S^1$ is a  spatial  circle  exhibit new 
dynamical scales  and new phenomena.  There are  chiral phase transitions 
taking place at $\Lambda^{-1}/N_c$ in the absence  of any change in center symmetry. 
 $\Lambda^{-1}/N_c$, invisible in (planar) perturbation theory,  
is also the scale where abelian versus non-abelian confinement regimes meet. 
  Large $N_c$ volume independence 
(a working Eguchi-Kawai reduction) provides new insights  and  independently 
confirms the existence of these scales. 
 We show that 
certain phases and scales are outside the reach of 
holographic (supergravity) modeling of QCD. 
\end{abstract}

\maketitle

\section{\label{sec:level1} Quantum phase transitions in QCD on  $\R^{2,1} \times S^1$}
Consider an asymptotically free,  confining QCD-like gauge theory with a  rank $N_c$ gauge group $G$ and   $n_f$ flavors of  massless fermions in a vector-like representation 
${\cal R}$ of $G$ (or a mixture of representations),  formulated on a   
space with one compact dimension.  Such theories possess a 
 strong dynamical  scale  $\Lambda$  as a consequence of dimensional transmutation. 

There are three  pieces of conventional wisdom associated with this class of  theories.  
{\bf a)}  Even though  the  chiral symmetry and  center symmetry are 
independent symmetries,  their  realization are  parametrically entangled.  
{\bf b)} If  phase transitions occur,  
they  must take place  in the numerical vicinity of strong scale $\Lambda$. 
{\bf c)}  If one takes the  large $N_c$ limit   (in the conventional 
 't  Hooft sense \cite{'t Hooft:1973jz}), the scale of the phase transitions must  be  $O(N_c^0)$.

Lattice gauge theory 
simulations  unambiguously  demonstrate that in finite temperature phase 
transitions,    the above conventional wisdom is  correct. 
Any  confining  QCD-like  theory, regardless of the representation  
${\cal R}$ of  the  fermions,   will undergo a center symmetry (or approximate center symmetry if 
center symmetry is absent)  changing 
confinement-deconfinement  transition  accompanied with the chiral symmetry 
transition which occurs around the strong scale  $\Lambda$   
\cite{ Aoki:2006we,Cheng:2007jq}. 
 The  fact that a very high temperature phase with broken  center symmetry  
cannot support any kind of chiral condensate can be proven 
rigorously \cite{Tomboulis:1984dd}.     The second assertion  seems to be 
robust due to the absence of any other dimensionful parameters  in the theory.  
The last one  states that the large $N_c$ limit should be  a good guide to 
probe such transitions.  This is also an underlying assumption in recent 
holographic models of QCD. 
The thermal QCD-like theories  provide full support 
for these three assertions.

In this letter, 
we wish to examine  the validity of these common 
assertions by testing them in a slightly 
different setup.  We wish  
to classify the phases of  {\it zero temperature} 
QCD-like theories on a space with one  compact spatial dimension,
  $\R^{2,1} \times S^1$, as a function of $S^1$ radius.  
We will work in a Euclidean setup, $\R^3 \times S^1$, 
hence   the anti-periodic or periodic  spin  structure of the fermions,  ${\cal S}^{\mp}$, 
 along the $S^1$ circle, 
determines whether $S^1$ is  thermal or spatial  circle, respectively. 

The rationale  behind this proposal  lies in the  sharp qualitative differences between  
 thermal and  quantum fluctuations.   In this sense, 
using  periodic 
boundary conditions for fermions is equally physical, and in some ways a  better guide to single 
out the quantum fluctuations.  The phase transitions   on spatial $S^1
\times \R^3$ (if any), are induced by zero temperature quantum mechanical 
fluctuations rather then the  thermal fluctuations.  Hence these 
are quantum phase transitions, as often appear in  
 condensed matter physics \cite{Sachdev}. 
  
These quantum phase transitions  reveal new and surprising phenomena in 
 clear contradiction  with the conventional wisdom.  Most notably, there are cases with chiral transitions 
in the absence of {\bf any} change in center symmetry realizations. More interestingly,
some such transitions takes place at a scale  parametrically split from 
$\Lambda^{-1}$ by factors of $N_c$. For example, in QCD(adj) with multiple fermions  and 
periodic boundary conditions (in which center symmetry never breaks \cite{Kovtun:2007py}), we will show that the chiral transition scale is 
 \begin{eqnarray}
 L_{\chi} = c \Lambda^{-1}/ N_c 
\label{quantum} 
\end{eqnarray} 
where $c$ is an order one numerical factor.  The existence  of the  scale  (\ref{quantum})
is  the main new result of   this letter.   This  scale is invisible in the    planar 
perturbation theory   which combines the   two coupling $(g^2, N_c)$ into a single  't  Hooft coupling $\lambda= g^2N_c$ \cite{'t Hooft:1973jz}.  The reason is, the planar  perturbative expansion does not respect the center symmetry of the small $S^1$ regime. 

The presence of such  a suppressed 
scale  may seem exotic. However, 
 I will argue that it   must exist in {\bf all} confining QCD-like theories in the following sense:  Were  the center symmetry  to remain unbroken in a multi-flavor QCD-like theory 
 (either thermally or spatially compactified),  
 a chiral transition would still take  place and it would occur at 
  $c \Lambda^{-1}/ N_c $ (and not at  $\Lambda^{-1}$).    Hence,   (\ref{quantum})
   is in fact the natural scale of the chiral transition in QCD, 
   answering a question raised in  \cite{Mocsy:2003qw}.

\subsection{Classification} The vector-like  gauge theories with massless fermions   formulated 
on  $\R^{3} \times S^1$
split into at least four  categories according to 
 their spatial center ($C_s$), and chiral ($\chi$)  symmetry
realizations. 
The classes  are:
\begin{itemize}
{\item [{\bf i)}] The theories in which $C_s$ and $\chi$ symmetry 
realizations are entangled, either into a single transition or a non-parametrically 
separable  double transition,  
and   both transitions occur   around strong scale $L_c \sim \Lambda^{-1}$.}
{\item[{\bf ii)}] The theories without any phase transitions}
{\item[{\bf iii)}] Center symmetric theories with a  chiral 
transition at a suppressed scale $L_{\chi} \sim \Lambda^{-1}/N_c$.  }   
{\item[{\bf iv)}] The theories in which $\chi$ and $C_s$ symmetries are entangled, and   both transitions  occur  around  a suppressed  scale $L_c \sim \Lambda^{-1}/N_c^k$, for some  $k >0$.}
\end{itemize}

\begin{figure}[ht]
\begin{center}
\includegraphics[angle=-90, width=0.4\textwidth]{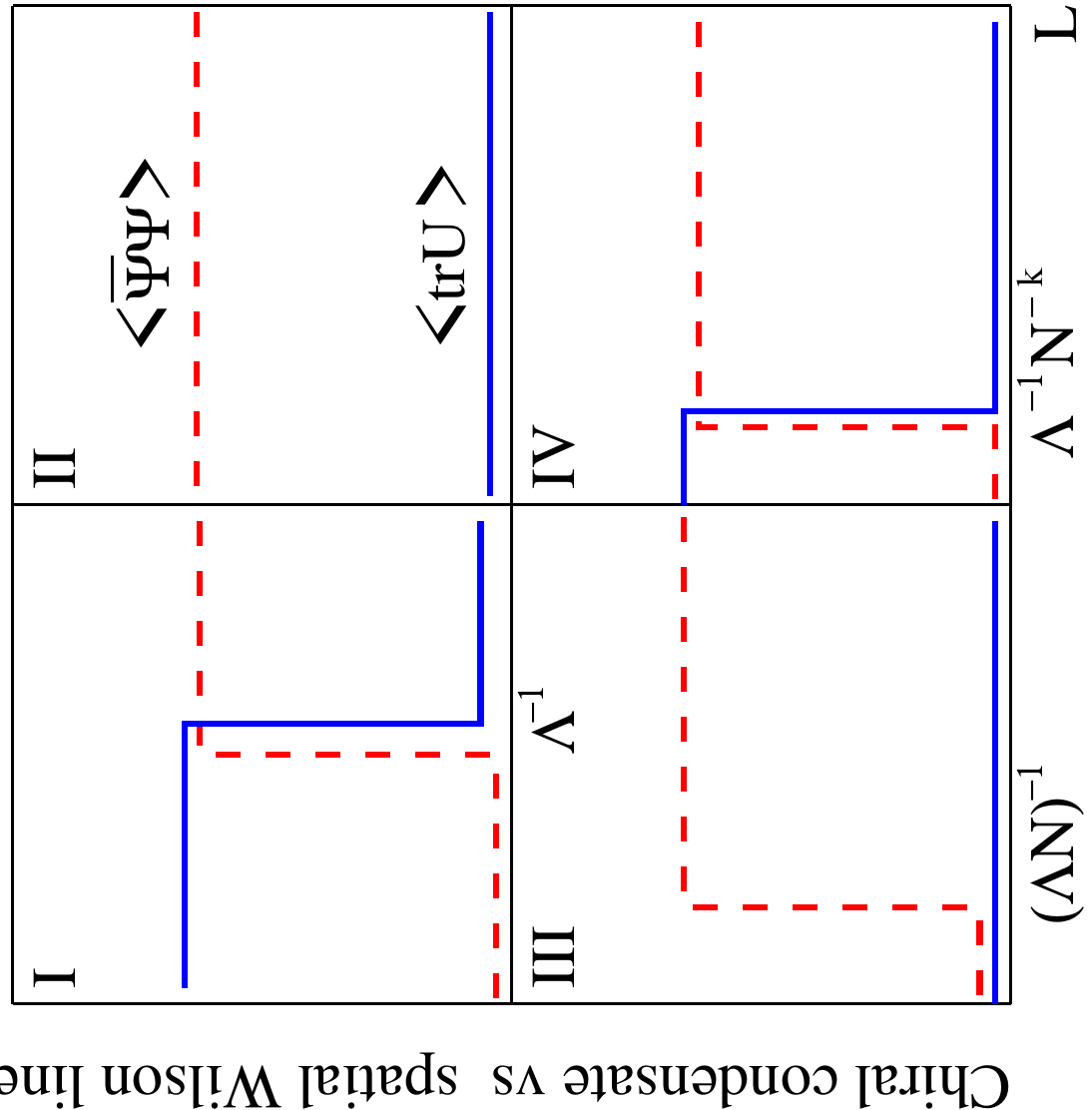}
\caption{The cartoons of the zero temperature center (blue, straight)  and chiral  (red, dotted) symmetry  realizations  in vector-like theories
with periodic boundary conditions for fermions.  In the thermal case, only  class I) 
is possible.}
  \label {fig:classes}
\end{center}
\end{figure}

As asserted earlier, this richness is   a reflection 
 of the distinction of  quantum versus thermal fluctuations. 
 Sufficiently  high temperatures melt  the hadrons and glueballs  into quarks and gluons 
 regardless of the fermionic  matter content of the  theory.  
The quantum fluctuations may or may not 
 generate quantum melting.  This depends on the matter content  of the  theory. Therefore, probing a theory solely  based on   quantum fluctuations may be a 
 better guide to study QCD dynamics.


{\bf A corollary of classification:}   At $N_c = \infty$,  
iii) and iv) degenerate  into  ii).  This means, the critical radius of 
transition for the $N_c= \infty$ theory is $L_c=0$.  

{\bf Phase transition as a function of $N_c$:} The corollary implies that, although an  $SU(3)$ gauge theory on small, fixed $S^1 \times \R^3$  may be in a confinement without chiral symmetry breaking phase, on the same  four-manifold, the $SU(\infty)$  theory   is in a chirally asymmetric, confinement phase.  In other words, 
the large $N_c $ limit
is separated from the small  $N_c$ theory on the same four-manifold  by a phase transition. 
In such circumstances, the large $N_c$ is not a good guide to study phases of the small $N_c$ 
theories, and such cases are generic.

\subsection{Applications}
We  assume the vector-like theories of our interest are in center symmetry unbroken, 
chiral symmetry broken phase at large $S^1 \times \R^3$.   
In the small $S^1$ regime,
the center symmetry realization can be determined by conventional techniques,  by  evaluating 
the one loop  potential in the background of the 
constant spatial  Wilson line $U( x ) = P e^{i\int A_4(x, x_4) dx_4}$,  and by using periodic boundary conditions for fermions.  
Below are the simplest examples for the four classes. 

{\bf i)} Consider QCD with $n_f$ complex  representation fermions, such as 
 fundamental, antisymmetric or symmetric.  In these theories, 
the examination of the  effective  potential  
$V_{\rm eff}[U(x)]$ on sufficiently 
small $S^1$  shows that 
$\langle \tr U(x) \rangle \neq 0$, and center symmetry is broken. The effect 
is  due to $O(N_c^2)$ part of the potential. 
The perturbative one loop 
potential also renders fermions massive with a gap of order  $\frac{1}{L}$. 
Hence, the long distance correlators of chiral operators decays exponentially, 
with no formation of any condensate. Although this class do not exhibit 
any novel scales, 
the center broken phase also breaks C, P, T as well as CPT. 
\cite{Unsal:2006pj}.
In this sense, it is an interesting phase. 
The lattice studies of such  QCD-like 
theories with one 
compact spatial dimension is  initiated recently  in Refs.~\cite{DeGrand:2006qb,Lucini:2007as}, and  for $SU(3)$ gauge theory, a CPT breaking  and restoring transition is  observed. 

{\bf ii)} The examples of this class are $\N=1$ SYM theory with arbitrary 
gauge group $G$.  The one loop potential is $V_{ \rm eff}[U]=0$    due to 
supersymmetry. Nonetheless,    
 a nonperturbatively generated potential provides a repulsive 
interaction among eigenvalues of the spatial Wilson line, and the spatial center symmetry 
$C_s$  is unbroken.  

The unbroken  $C_s$ in the {\it weakly} coupled regime implies 
gauge symmetry ``breaking"
down  to the maximal abelian subgroup, $G \rightarrow {\bf  Ab}(G)$ at large distances. 
Consequently, 
the fermion modes along the Cartan subalgebra of $G$ remains massless, and 
they are part of the long distance physics.  The fractional instanton (monopole) effects 
are  sufficient to produce a chiral condensate and spontaneous breaking 
of discrete chiral symmetry. 
The IR dynamics in this 
case is drastically different from both case  i) and the 
thermal compactification.  
Small $S^1$ exhibits abelian  confinement  and discrete chiral symmetry breaking 
 \cite{Katz:1996th,Davies:2000nw,Unsal:2007jx}.

{\bf iii)}  The  asymptotically free  QCD-like theories  with $n_f>1$ adjoint 
Majorana fermion  are  in this category.   
 Introducing adjoint fermions with  periodic boundary conditions ${\cal S^{+}}$
 stabilizes the center symmetry breaking instability \cite{Kovtun:2007py}.  
 The one loop potential for QCD(adj) with 
${\cal S^{+}}$ is proportional to the negation of the one of pure YM theory: 
\begin{equation}
V_{\rm eff, +}^{\rm QCD(adj)}[U(x)] = ( 1-n_f) V_{\rm eff}^{\rm YM}[U(x)] 
\label{Eq:Potadj}
\end{equation}
This implies unbroken spatial center symmetry in QCD(adj) with ${\cal S^{+}}$ 
at small $S^1$  weak coupling regime,   and consequent dynamical abelianization 
    down to ${\bf  Ab}(G)$  at large distances.  QCD(adj)   has a
continuous    $SU(n_f)$  chiral symmetry.  At small $S^1$, the theory exhibits confinement 
without chiral symmetry breaking  as shown in \cite{Unsal:2007jx} for $N_c=$
 few. At   long distances, abelian confinement is operative just like 
 the $\N=1$ SYM theory in the same regime, and the
  photons acquire mass via  magnetic bion mechanism.  
   At length scales larger than the inverse photon mass, the long distance effective theory reduce to a NJL-type Lagrangian  of zero mode fermions which is known to possess two phases: a weak coupling unbroken phase and a strong coupling broken phase.     
 Although the strong coupling phase is outside the region of validity of the  effective theory, the transition is taking place  just at its  boundary where $e^{-S_0} \sim 1 $. Here, $S_0 =  \frac{8 \pi^2} {g^2(m_W)  N_c}$ is the monopole (fractional instanton) action.

 The   interesting aspect,   which is not addressed in    \cite{Unsal:2007jx} in generality,  is that 
  in  terms of compactification radius $L$, $N_c$ and $\Lambda$, this translates 
 into   $ L  N_c \Lambda  \sim 1$.  This is also the scale where the non-perturbatively generated photon mass becomes equal to the lightest $W$-boson mass $m_W=2\pi/(N_cL)$  and 
 a long distance description based on 
  ${\bf  Ab}(G)$ ceases its validity.  The theory moves from an the abelian confinement 
  to non-abelian confinement regime \cite{Shifman:2007rc}. In multi-flavor QCD(adj) theory, this is associated with a solo chiral phase transition.

There is a deeper reason for the existence of a suppressed  chiral transition 
scale  $\Lambda^{-1}/N_c$.   The non-perturbative physics of large $N_c$  QCD-like 
gauge theory is independent of the volume of the $S^1$ so long as the center symmetry is 
unbroken \cite{Eguchi:1982nm, Yaffe:1981vf,Bhanot:1982sh}. This is the  volume independence  property at large $N_c$, also known as 
  Eguchi-Kawai reduction. For pure YM, a full  EK reduction fails just because the center symmetry   breaks spontaneously. 
   QCD(adj) with ${\cal S}^{+}$ is the first continuum gauge theory example  
    which satisfies the volume independence all the way down to arbitrarily small volumes 
    \cite{Kovtun:2007py}, as opposed to the partial reduction discussed in \cite{Narayanan:2003fc} 
    and other schemes which are recently shown to fail \cite{Teper:2006sp, Azeyanagi:2007su, Bringoltz:2008av}.
    
       The  independence of the physics from the $S^1$ size at $N_c=\infty$ limit   implies that the chiral symmetry realization and chiral condensate must be independent of $S^1$ 
  radius in QCD(adj).   Therefore, whatever chiral transition takes place at finite $N_c$ and finite 
  $S^1$  must be pushed to $L_\chi=0$ at $N_c=\infty$. 
Also note that   the existence of  ${\bf  Ab}(G)$  regime \cite{Unsal:2007jx}  and the large $N_c$ volume independence   \cite{Kovtun:2007py} are consistent with each other, because the region of validity of  ${\bf  Ab}(G)$  regime 
  is pushed into an arbitrarily narrow sliver as $N_c$ is taken large. 

{\bf iv)} The YM theory with one  adjoint   Majorana and $n_f \geq 1$   
fundamental Dirac fermions all with ${\cal S}^{+}$ is in this class.  For $n_f=0$, the theory is 
 just $\N=1$  SYM.  
  Due to supersymmetry of $n_f=0$ background, $V_{\rm eff}[U]$ is 
  $0  \times O(N_c^2) + n_f O(N_c)$ where the first non-vanishing contribution is due to  $n_f$  fundamental 
fermions. The minimum of the $V_{\rm eff}[U]$ is at  $U=-1$ and the center is broken at small $S^1$. 
The interesting behavior of this theory is due to the fact that  the  one loop potential 
 is $n_f O(N_c)$.  Recall from thermal QCD --the case in which thermal one-loop potential is 
 $O(N_c^2)$--  that the usual electric mass of the holonomy  
 (or the $A_4$  field)  is $m_{e} \sim  \frac{\sqrt \lambda}{L}$.   In our case, 
the mass of the   $A_4$  field is  anomalously small  \cite{Ofer}
\begin{equation}
m^*= \frac{\sqrt \lambda}{L}  \sqrt \frac{n_f}{N_c} \equiv m_{e} \sqrt \frac{n_f}{N_c} 
\end{equation}
In the large $N_c$ limit, $m^*/m_e \rightarrow 0$ and the one loop effective potential seen by the $N_c$ eigenvalues becomes arbitrarily flat.  A classical moduli space opens up at this level of analysis. 
As  in the $\N=1$ SYM case, the non-perturbative potential restores the center symmetry. This means, the chiral and center symmetry transition scales must be suppressed scales relative to $\Lambda^{-1}$. 


Another way to realize that there must exist a  suppressed  chiral and center symmetry transition scale  is to use  the large $N_c$ volume independence.   
In the $N_c=\infty$ limit, the SYM theory obeys  volume 
independence. The theory on $\R^4$ is nonperturbatively equivalent to the 
theory on $\R^3 \times S^1$ for any finite $S^1$.  In other words, 
(somehow counter-intuitively), the    $N_c=\infty$ limit of $\N=1$  SYM lacks a weak coupling 
${\bf Ab}(G)$ description at long distances regardless of how small $S^1$ is, so long as it is finite. 
 The addition of fundamental fermions on small $S^1 \times \R^3$  is same as adding fundamental fermions to the theory on $\R^4$. 
Consequently, at $N_c=\infty$ limit, fundamental fermions cannot induce a center symmetry breaking in this theory.  


\subsection{Implications and comments}

{\bf Refined (abelian) large $N_c$ limit:} 
For any confining QCD-like gauge theory which remains center symmetric at arbitrarily small $S^1 \times \R^3$, there exists  a double-scaled,   refined large $N_c$ limit.  
In this limit,   
\begin{equation}
\Lambda^{-1} = O(N_c^0), \; \;\; \;  LN_c\Lambda = O(N_c^0)   \ll 1 
\end{equation} 
are   held fixed. 
 The short distance is 
$U(\infty)$ and 
long distance gauge structure is a $[U(1)]^{\infty}$ mimicking the ${\bf Ab}(G)$ 
structure of the small $N_c$, small $S^1$ center symmetric theories.   
The existence of an abelian large $N_c$  limit was first shown in non-vector-like 
${\cal N}=2$ SYM theory  \cite{Douglas:1995nw}. Realizing that such a limit exists 
and   is generic in QCD-like theories is new.  We expect the refined large $N_c$ limits to be  generically solvable as in \cite{Unsal:2008ch}.

{\bf A no-go theorem for  holographic  (supergravity) models of QCD:}  
Since supergravity is the classical $N_c=\infty$ limit, the small radius phases of the classes iii) and iv) are invisible in the holographic modeling of QCD as a result of  corollary.   
 In particular,  the  confinement without chiral symmetry breaking phase and the associated solo chiral transitions are outside the reach of  supergravity approximation. 
A stringy  improvement of holographic  models, which   incorporates the $O(1/N_c)$  effects,  is needed to  find the  phase transitions in classes iii) and iv).  

Confirming this result, the  holographic models of QCD so far did not find any 
confinement without chiral symmetry breaking phase, although chirally asymmetric deconfined phases were found \cite{Aharony:2006da,Parnachev:2006dn}. This non-observation can naturally be explained  by large $N_c$ volume independence.  
It  is also tied with ``dynamical abelian dominance" of the refined large $N_c$ limit, 
 one  cannot describe phases with $O(N_c)$ weakly coupled fields as in  \cite{Douglas:1995nw, Unsal:2008ch, Shifman:2008ja}
in supergravity approximation \cite{Ofer}.   


{\bf Testing on  the lattice:} One may   wonder why these novel scales  are not already seen in lattice gauge theory simulations. The reason  is simple.  So far, there exist no non-thermal  lattice simulation for classes iii) and iv).  
 
Even in the thermal setting,  the confinement without chiral symmetry breaking phase can be 
observed as follows: Assume  the center symmetry is stabilized   at small $S^1$ by a 
double trace deformation \cite{Shifman:2008ja,Unsal:2008ch}.  
In   multi-flavor theories, there will still be 
a solo chiral transition from a confined chirally asymmetric to a confined chirally symmetric phase, which will take place at  $\beta_\chi= c \Lambda^{-1}/N_c$ due to similar reasons as 
in QCD(adj). 
 In this sense,  the  natural scale of chiral transition   is  again 
 (\ref{quantum}). 
 In the  thermal QCD, this  scale is  shadowed by the 
 deconfinement transition which occurs  around $\Lambda^{-1}$ and forces 
 the chiral symmetry to restore  \cite{Tomboulis:1984dd} far before $\Lambda^{-1}/N_c$.
  The simulation of deformed QCD  and the quantum phase transitions  
  are   feasible by conventional techniques. 

{\bf Summary:} In any confining QCD-like theory which always remains center symmetric 
(either by non-thermal fluctuations  or by deformations) on $\R^3 \times S^1$,  the scale    
$  \Lambda^{-1}/N_c$ is   the most important scale. 
For smaller radius, the long distance theory abelianizes down to ${\bf Ab}(G)$ and abelian confinement is operative. For larger radius, non-abelian confinement is valid. For zero and one flavor theories, the abelian and non-abelian confinement regimes are smoothly connected  \cite{Shifman:2008ja,Unsal:2008ch}.  
For multi-flavor theories, there must exist a single  chiral transition 
from  a chirally symmetric abelian confinement regime to a chirally asymmetric nonabelian 
confinement regime occurring at 
 $c \Lambda^{-1}/N_c$.  
Since the non-abelian confinement and volume independence 
 holds for $LN_c\Lambda >1$,  and the theory just above this critical radius is equivalent to QCD 
   on $\R^4$ up to $O(1/N_c^2)$ corrections,    understanding the dynamics of QCD in the  vicinity of (\ref{quantum})  may hold the necessary insights into a fuller  
 understanding of  the theory. 

{\bf Acknowledgments:}
I am grateful to  O.Aharony, R. Brower, D. Harlow,  D.Kutasov, A. Parnachev, M. Shifman and L. Yaffe 
for useful  discussions. 
This  work  is  supported by the U.S.\ Department of Energy Grant DE-AC02-76SF00515.


\end{document}